\documentclass{aa}  
\usepackage{graphicx}
\catcode`\@=11
\def\gtsima{\ifmmode{\mathrel{\mathpalette\@versim>}}
    \else{$\mathrel{\mathpalette\@versim>}$}\fi}
\def\ltsima{\ifmmode{\mathrel{\mathpalette\@versim<}}
    \else{$\mathrel{\mathpalette\@versim<}$}\fi}
\def\@versim#1#2{\lower 2.9truept \vbox{\baselineskip 0pt \lineskip 
    0.5truept \ialign{$\m@th#1\hfil##\hfil$\crcr#2\crcr\sim\crcr}}}
\catcode`\@=12
\usepackage{txfonts}
\begin{document}
   \title{Mass downsizing and ``top-down'' assembly of early-type galaxies}

   \author{A. Cimatti \inst{1}
          \and
          E. Daddi \inst{2}
          \and
	  A. Renzini \inst{3}
          }

   \offprints{A. Cimatti}

\institute{INAF - Osservatorio Astrofisico di Arcetri, Largo E. Fermi
5, I-50125 Firenze, Italy \\
\email{cimatti@arcetri.astro.it}
\and
{\em Spitzer} Fellow; National Optical Astronomy Observatory, 950 N. Cherry Ave., Tucson, AZ, USA \\ 
\email{edaddi@noao.edu}
\and
INAF - Osservatorio Astronomico di Padova, Vicolo
dell'Osservatorio, 5, I-35122 Padova, Italy \\
\email{arenzini@pd.astro.it}
}

\date{Received .....; accepted .....}

\abstract
{} 
% aims heading (mandatory)
{We present a new analysis of the rest-frame $B$-band COMBO-17 and DEEP2
luminosity functions (LFs) of early-type galaxies (ETGs) as a function of
luminosity and mass. Our aim is to place new stringent constraints on the
evolution of ETGs since $z\sim1$.}
% methods heading (mandatory)
{We correct the LF($z$) data for the luminosity dimming assuming pure 
luminosity evolution. However, instead of relying on stellar population 
synthesis model--dependent assumptions, we adopt the empirical luminosity 
dimming rate derived from the evolution of the Fundamental Plane of 
field and cluster massive ETGs.}
% results heading (mandatory)
{Our results show that the amount of evolution for the ETG 
population depends critically on the range of luminosity and masses
considered. While the number density of luminous (massive) 
ETGs with $M_{\rm B}($z=0$) < -20.5$ (${\cal M}>10^{11}$ 
M$_{\odot}$) is nearly constant since $z\sim 0.8$, less luminous galaxies 
display a deficit which grows with redshift and that can be explained with 
a gradual population of the ETG ``red sequence'' by the 
progressive quenching of star formation in galaxies less massive than 
$\sim 10^{11} M_{\odot}$. At each redshift there is a critical mass above
which virtually all ETGs appear to be in place, and this fits well in the 
now popular ``downsizing'' scenario. However, ``downsizing'' does not 
appear to be limited to star formation, but the concept may have to be 
extended to the mass assembly itself as the build-up of the most
massive galaxies preceeds that of the less massive ones. This
evolutionary trend is not reproduced by the most recent theoretical 
simulations even when they successfully reproduce ``downsizing'' in 
star formation.}
{}


\authorrunning{A. Cimatti et al.}
\titlerunning{Anti-hierarchical assembly of ETG}

\maketitle
%
%________________________________________________________________

\section{Introduction}

Early-type galaxies (ETGs) dominate the top--end of the local ($z\sim0$)
galaxy mass function, and therefore are crucial probes to investigate the 
history of galaxy mass assembly. There is now convincing evidence 
that cluster ETGs formed the bulk of their stars very early in 
the evolution of the Universe, i.e., at $z\gtsima 3$, while ETGs 
in low density (``field'') environments did so 1 or 2 Gyr later, 
i.e., at $z\gtsima 1.5-2$ (e.g., Thomas et al. 2005, Renzini 2006,
and refs. therein ).
However, in the current hierarchical scenario star formation and mass
assembly are not necessarily concomitant processes in galaxy
formation: stars may well have formed at very high redshift in
relatively small units, but only at lower redshift (e.g., $z \ltsima 1$)
they may have merged together to build the massive ETGs that we see in the
nearby universe. 
%If so, one would expect the comoving number density
%of ETGs to decrease with redshift, and the mass assembly process can
%be directly investigated by mapping how the luminosity and mass
%functions of ETGs evolve with redshift.
Only recently, with the advent of wide--field surveys (i.e. $\gtsima 
1$ square degree) it became possible to reduce the strong cosmic variance 
which affects ETG studies, and place more stringent
constraints on the ETG evolution. Yet, conclusions reached by
different studies still appear quite discrepant with respect to each
other. On the one hand, some results indicate little evolution.  
For instance, the VIMOS VLT Deep Survey (VVDS) shows that the rest-frame 
$B$-band LF of ETGs with $I<24$ is consistent with passive evolution up 
to $z\sim1.1$, and the number of bright ETGs decreases by only $\sim$40\% 
from $z\sim0.3$ to $z\sim1.1$ (Zucca et al. 2006). Similarly, the 
Subaru/XMM-Newton Deep Survey (SXDS) selected a large sample of ETGs at 
$z\sim1$ down to $z^{\prime}<25$, finding their number density at 
$M_{\rm B}^*$ to be up to 85\% that of ETGs at $z=0$ (Yamada et al. 2005).
Moving ETG selection to even longer wavelengths, $K$-band selected surveys
uncovered a substantial population of old (1-4 Gyr), massive (${\cal
M}>10^{11} M_{\odot}$), passively evolving E/S0 galaxies at $1<z<2$
(e.g. Cimatti et al. 2002a, 2004; McCarthy et al. 2004; Glazebrook et al. 
2004: Daddi et al. 2005a; Saracco et al. 2005), and showed that the 
high-luminosity/high-mass tails of the total luminosity and stellar mass 
functions (which are dominated by old ETGs) evolve only weakly to 
$z\sim$0.8--1 (e.g. Pozzetti et al. 2003; Fontana et al. 2004; Drory et 
al. 2005; Caputi et al. 2006; Bundy et al. 2006). 

On the other hand, the COMBO-17 (Bell et al. 2004a) and DEEP2 
(Faber et al. 2005) surveys indicate a stronger evolution of the ETG 
population characterized by a faster decrease of the number density with
redshift. These two surveys rely on either high-quality photometric 
(COMBO-17) or spectroscopic redshifts (DEEP2), and ETGs were selected 
in the optical ($R<24$) for lying on the ``red sequence'' in
the rest-frame $U-V$ vs $M_{\rm B}$ color-magnitude relation, with as
many as $\sim 85\%$ of them being also morphologically early-type
(Bell et al. 2004b). The red sequence of ETGs in COMBO-17 becomes
progressively bluer in the rest frame $U-V$ color, going from $z=0$ to
$z=1$, consistent with pure passive evolution of stellar populations
formed at high redshift ($2<z_f<5$, Bell et al. 2004a). The
rest-frame $B$-band luminosity function in the various
redshift bins was 
fitted (both in COMBO-17 and DEEP2) with a Schechter
function, thus 
determining $M_{\rm B}^*(z)$ and $\phi^*(z)$, having
assumed the faint-end 
slope of the LF to be independent of $z$, and
fixed at its low-redshift 
value ($-0.5$ and $-0.6$, respectively for
DEEP2 and COMBO-17). As such, 
this latter assumption virtually
excludes ``downsizing'' in galaxy 
formation, i.e. star formation
ending first in massive galaxies than in 
lower mass ones, a trend for
which ample evidence is growing both at 
low and high redshift (Cowie
et al. 1996; Thomas et al. 2005; Tanaka et 
al. 2005; Kodama et
al. 2004; van der Wel et al. 2005; Treu et al. 2005; 
di Serego et
al. 2005; Juneau et al. 2005; Feulner et al. 2005; Bundy 
et al. 2006).  With these
assumptions, Faber et al. (2005) analyzed the 
DEEP2 data and
re-analyzed the COMBO-17 data as well, finding $L_{\rm B}^*$ 
to
increase by a factor $\sim 1.5$ ($\sim 2.4$) between $z=0.3$ and 1.1,
while $\phi^*$ drops by a factor $\sim 2.5$ ($\sim 4$), where values
in parenthesis refer to COMBO-17 data. In the case of COMBO-17 data, 
also the brightening of $L_{\rm B}^*$ is consistent with passive 
evolution. In both cases the $B$-band
luminosity density provided by 
ETGs ($j_{\rm B}\propto L_{\rm
B}^*\phi^*$) remains flat up to $z\sim 
0.9$, while pure passive
evolution would have predicted an increase of 
the luminosity density
by a factor $\sim$2--3. Thus, while the ETG colors 
and $M_{\rm
B}^*(z)$ follow the passive evolution,  the number
density of ETGs does not, both Bell et al. (2004a) and Faber et
al. (2005) concluded that the stellar mass in red sequence ETGs has
nearly doubled since $z\sim 1$. Bell et al. further interpreted this
result as strong evidence in supports of the semi-analytic model of
galaxy formation by Cole et al. (2000), which indeed predicts $j_{\rm
B}$ to remain constant to $z\sim 1$. Both Bell et al. and Faber et al.
then argue for a major role of ETG-ETG merging (now called "dry"
merging) in the build up of the ETG population between $z\sim 1$ and
$z \sim 0$, especially for the most massive ones given the shortage of
massive star-forming precursors. 

In this paper we attempt at offering a new and consistent interpretation of
the COMBO-17 and DEEP2 data by exploring the dependence of ETG evolution
on luminosity and mass, and where the passive evolution and number
density decline with redshift are reconciled in a scenario that is
also consistent with all other empirical evidences on ETGs, at low, as
well as high redshift. We adopt $H_0=70$ km s$^{-1}$ Mpc$^{-1}$,
$\Omega_{\rm m}=0.3$ and $\Omega_{\Lambda}=0.7$ and give magnitudes in
Vega photometric system.

\section{Lost and found progenitors to local ETGs}

In this section, we analyze with a new approach the COMBO-17 and DEEP2
LFs in order to investigate the evolutionary link between the high-$z$
ETGs and their local descendants, as well as the metamorphosis of
other kinds of galaxies to progressively turn passive and qualify as
ETGs. 
In doing so, we assume that ETGs which are passive at high redshift
will remain passive through $z=0$ and will be subject to pure luminosity 
evolution. This is justified by the color evolution of the red sequence 
following the expectations for pure passive evolution (Bell et al. 2004a).
Under this assumption, we {\it evolve} down to $z=0$ the ETG LF($z$) in
each redshift bin, and compare it with the corresponding LF of local ETGs. 
Instead of relying on stellar population synthesis models, we adopt the 
passive luminosity dimming derived empirically from the evolution of the 
Fundamental Plane for cluster ETGs, where $\Delta$log$({\cal M} / L_{\rm B}) 
= (-0.46 \pm  0.04)z$ (van Dokkum \& Stanford 2003; Treu et al. 2005; di 
Serego Alighieri et al. 2005). Massive field ETGs appear to follow the same
relation, whereas less massive ones evolve faster in luminosity,
indicative of younger ages (Treu et al. 2005; di Serego Alighieri et
al. 2005). We first apply this relation to all ETGs, independent of their 
luminosity, and then we comment on the effect of a luminosity-dependent 
rate of evolution. In this way, the luminosity of high-$z$  ETGs
is decreased by $\Delta$log$L_{\rm B}= 0.46\, z$ to
obtain the luminosity to which they would fade by $z=0$ due to their 
passive evolution, i.e., their $M_{\rm B}$-band magnitude at $z=0$ is 
obtained as:

$$M_{\rm B}(0)=M_{\rm B}(z)+(0.46 \times 2.5 \times z)=M_{\rm B}(z)+1.15 \ z$$

This evolutionary rate is consistent with that predicted by stellar
population synthesis models for stars formed at $z \gtsima 1.5$ (Bell
et al. 2004a).
For each redshift bin in the range $0.25<z<1.10$, the resulting {\it
evolved} LF to $z=0$ is then compared to the data points of the
local ($z=0$) LF derived from Bell et al. (2004a) based on SDSS
data. The result is shown in Fig. 1, where besides to COMBO-17 and
DEEP2 galaxies, the same procedure has been applied also to the SXDS
sample of ETGs (Yamada et al. 2005).

\begin{figure}
\centering
\includegraphics[width=9cm]{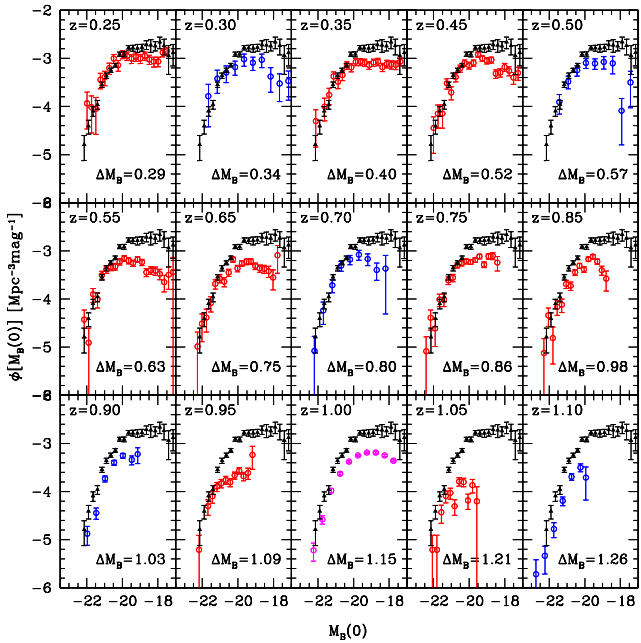}
\caption{
The $B$-band LF of ETGs from COMBO-17 (Bell et al. 2004a, red), 
DEEP2 (Faber et al. 2005, blue), and SXDS (Yamada et al.
2005, magenta) ``evolved'' to $z=0$ and compared to
the local LF derived by Bell et al. (2004a) from SDSS
data (black filled triangles). The amount of applied
passive evolution ($\Delta M_{\rm B}$) is also indicated. 
}
\label{}
\end{figure}

In each panel, the amount of luminosity evolution ($\Delta M_{\rm B}$)
is shown, and the error bars in the number density $\phi$ include also
the cosmic variance contribution as estimated by Bell et al. (2004a)
and C. Willmer (private communication) for COMBO-17 and DEEP2,
respectively. It appears very clearly from Fig. 1 that the number
density of the ETGs populating the bright-end of the LF shows basically 
no evolution with redshift up to $z \sim 0.8$, whereas less luminous ETGs 
display a deficit which grows with redshift. 

Fig. 1 also shows that at any redshift there is a characteristic
luminosity above which the bright end of the passively evolved ETG LF
is consistent with the local LF, and the luminosity at which it
departs from it depends on redshift, becoming more luminous
for increasing redshifts. Fig. 2 displays this trend by showing the
absolute magnitude, $M_{\rm B}(0)$(60\%), at which in each redshift
bin of Fig. 1 the evolved LF is 0.6 times less than the local LF,
corresponding to the luminosity below which less than 60\% of the ETGs
are already in place. Fig. 2 clearly shows a trend of $M_{\rm
B}(0)$(60\%) becoming brighter for increasing redshifts, indicating
once more that the most massive galaxies are the first to reach the
ETG sequence, while less massive ones join it later, i.e., at lower
redshifts.  The trend shown in Fig. 2 is consistent with that of the
``quenching'' mass with redshift ${\cal M}_{\rm Q} \propto
(1+z)^{4.5}$ (Bundy et al. 2006), where most galaxies more
massive than ${\cal M}_{\rm Q}(z)$ have already turned passive.

\begin{figure}
\centering
\includegraphics[width=9cm]{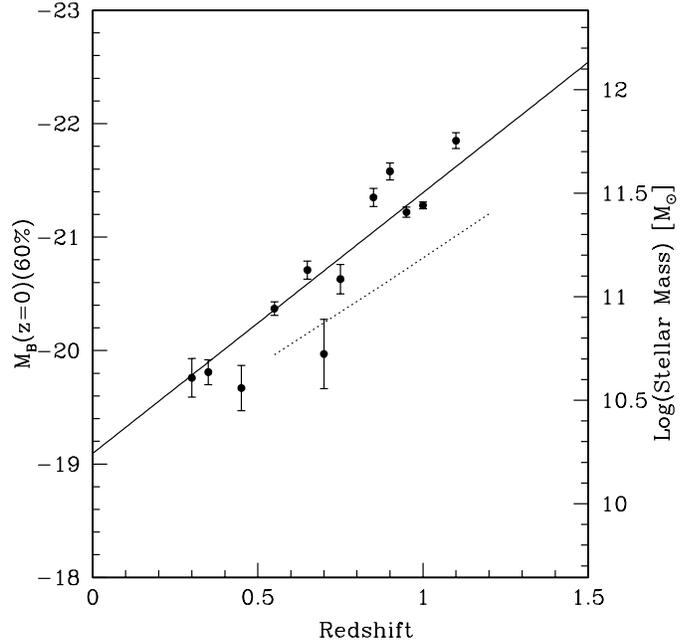}
\caption{
The luminosity (or equivalent mass) at which about 60\% of the ETGs 
are already in place. The dashed line represent the evolution of 
the ``quenching'' mass of Bundy et al. (2006).
}
\label{}
\end{figure}

This luminosity-/mass-dependent evolution of ETGs is further
highlighted in Fig. 3 which shows the evolution of the ratio\footnote{
$F(z)=\int_{M_1}^{M_2} \phi[M_{\rm B}(0),z]dM_{\rm B}(0)/ \int_{M_1}^{M_2} 
\Phi[M_{\rm B}(0)]dM_{\rm B}(0)$, where $M_1$ and $M_2$ are the 
integration extremes, 
$\phi$ are the LF data points, and $\Phi$ is the best--fit 
Schechter function of the LF($z=0$) (Bell et al. 2004a).} between the $z>0$ 
and $z\sim0$ ETG number density, splitted into the 
bright ($M_{\rm B}(0) < -20.5$) and the faint sample ($-20.5 < M_{\rm B}(0)
<-18.5$). The luminosity cut of $M_{\rm B}(0)= -20.5$ corresponds to
${\cal M} \sim 10^{11} M_{\odot}$ (di Serego Alighieri et al. 2005,
see their Fig. 13, from which we derive $M_{\rm B}(0)\simeq -1.825
\times$log$[{\cal M}/M_{\odot}]-0.4$). 
In order to avoid the mismatch between the data points of the 
LF($z\sim0$) and LF($z$), we use here the Schechter function derived by 
Bell et al. (2004a, Appendix) for the LF($z\sim0$).
F$_{\rm lum}(z)$ and F$_{\rm faint}(z)$ can be seen as the fractions
of assembled ETGs in the two luminosity (mass) intervals as a function 
of redshift. Fig. 3
shows that (within the error bars) the number density of ETGs with
${\cal M}>10^{11} M_{\odot}$ is nearly constant within the range 
$0<z<0.8$, and starts to decline significanlty for $z>0.8$. The values of
F$_{\rm lum}(z)$ at $z=0.25$ and $z=0.3$ are higher than the local one 
by 1-2$\sigma$, presumably as a result of a fluctuation due to the small 
volumes sampled at low-$z$. Instead, the number density of ETGs in the
fainter sample (or lower masses) shows a smooth decrease from $z\sim0$
to $z\sim0.8$ and beyond. Note that beyond $z \sim 0.9$ the COMBO-17 and 
DEEP2 LFs do not reach $M_{\rm B}(0)=-18.5$ (see Fig. 1), and therefore 
$F_{\rm faint}$ is relative to the integration of the LF($z$) down to the 
faintest $M_{\rm B}(0)$ available in the data points. Thus, most of the 
number density evolution of ETGs is confined to the galaxies fainter than
$M_{\rm B}(0) < -20.5$ (${\cal M} \ltsima 10^{11} M_{\odot}$). In other
words, up to $z\sim 0.8$ there is a sufficient number of massive ETGs
that will passively evolve to the local massive ETG and will
match their number density.
We tested if all the above results (Fig. 1--3) are solid with respect to
the adopted ingredients of our analysis. First, we verified that
the $\sim 10\%$ uncertainty in the adopted relation $\Delta$log$({\cal M}/
L_{\rm B})=(-0.46 \pm 0.04)z$ results in tiny shifts of the LFs, i.e., 
$\pm$0.025 mag at $z=0.25$ and  $\pm 0.1$ mag at $z=1$, and do not
affect the results. Second, our choice to dim the luminosity functions 
with a rate independent of luminosity is a {\it conservative} one: adopting a
luminosity-dependent rate of evolution consistent with the results of
Treu et al. (2005) and di Serego Alighieri et al. (2005) would have
made the number density deficit of faint ETGs even more marked. Third, 
we also tested that the results do not change significantly if instead
of the Bell et al. (2004a) best fit, we use an interpolation which 
provides a closer match to the local LF.

\begin{figure}
\centering
\includegraphics[width=9cm]{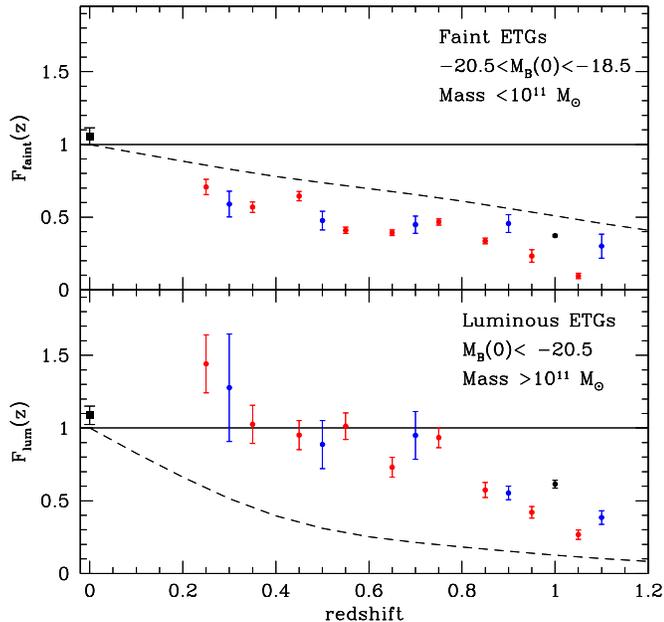}
\caption{
The evolution of the fractional number density of low-- (top) and 
high--luminosity (bottom) ETGs compared with the $z=0$ best--fit 
Schechter function of Bell et al. (2004a). The points at $z=0$
are not exactly equal to 1 due to the slight mismatch between the 
Schechter fit ($\Phi$) and the actual SDSS data points ($\phi$). 
Red, blue, black symbols at $z>0$ refer to COMBO-17, DEEP2 
and Yamada et al. (2005) respectively. The dashed lines correspond to 
the evolving fraction of the $z=0$ ETGs which have assembled 80\% of 
their stellar mass (from Fig. 5, bottom panel, of De Lucia et al. 2006).
}
\label{}
\end{figure}

\subsection{Incompleteness effects}

The drop in the space density of ETGs with increasing redshift being
stronger at faint luminosities suggested that it may at least in part
be due to sample incompleteness, thus exaggerating an apparent
downsizing effect.  Indeed, incompleteness may especially affect
optically selected samples, due to the large k corrections of ETGs.
Instead, $K$-band selected samples are less affected by this problem
having much smaller k-correction.  In order to assess the relevance of
this effect, we performed a simple test with the highly complete K20
sample (Cimatti et al. 2002b; Mignoli et al. 2005) by deriving the
fraction of the K20 spectroscopically-classified ETGs fainter than the
nominal limiting magnitude of the COMBO-17 and DEEP2 surveys (i.e., $R_{\rm
lim}\simeq 24$).  This fraction is $\sim$16-21\% and $\sim$50\% at
$0.8<z<1.0$ and $1.0<z<1.2$, respectively, showing that in the redshift
range  $0.8<z<1.2$ these surveys are 
affected by non-negligible incompleteness effects, which become very
important at $z \geq 1$. The K20 ETGs with $R>R_{\rm lim}$ have
absolute magnitudes at $z=0$ in the range of $-19.5< M_{\rm B}(0)<-18.5$
and stellar masses in the range of $10^{10}\ltsima {\cal M} < 3 \times 10^{10}\;
M_{\odot}$.  This test indicates that the highest redshift bins of the
COMBO-17 and DEEP2 surveys are affected by non-negligible
incompleteness, with up to $\sim$50\% of the $-19.5<M_{\rm B}
(0)<-18.5$ ETGs being missed at $z\sim1$. However, for objects with
$M_{\rm B}(0)<-19.5$ the COMBO-17 and DEEP2 samples should be 
fairly complete up to $z\sim 1$, hence genuine evolution appears to be
responsible for the observed drop of the LF for $-20.5<M_{\rm
B}(0)<-19.5$ at $z>0.8$.

\section{Main implications on ETG evolution}

The present new analysis of the COMBO-17/DEEP2 LFs shows that 
the evolution of ETGs, as inferred only in terms of the Schechter 
function best fit parameters $\phi^*(z)$ and $M^*(z)$ does not fully 
exploit all the available information because the evolution 
is a strong function of luminosity and mass. Thus, the redshift
evolution of the ETG number density must depend on the luminosity
(mass) range of each specific sample under consideration. The main 
implications of our analysis on the ETG evolution can be 
summarized as follows.

$\bullet$ When the empirical passive evolution dimming is used to
project the passive evolution of high redshift ETGs down to $z=0$,
then the high-luminosity end ($M_{\rm B}(0)<-20.5$) of the resulting
LF comes to a perfect match to the $z\sim0$ LF of local red-sequence
ETGs (Fig. 1). This implies that the number density of massive ETGs
(${\cal M} > 10^{11} M_{\odot}$) is nearly constant from $z=0.8$ to
$\sim 0$ (Fig. 3). 
This near-constant number density
of massive ETGs cannot be the result of a sizable fraction of them
leaving the red sequence (because e.g., of re-activated SF), while being
compensated by a near equal number joining the top end of the LF
by SF quenching. Indeed, as emphasized by Bell et al. (2004a) at every
redshift in the COMBO-17 sample the top end of the LF is dominated by
ETGs on the red sequence, with a comparatively insignificant number of
blue galaxies. Therefore, there is no room for an appreciable fraction of
red sequence galaxies to turn blue. 
Moreover, one could imagine a scenario in which dry merging operates in 
such a way to keep constant the number of massive ETGs (e.g., ${\cal M} > 
10^{11} M_{\odot}$). However, as dry merging would inevitably increase the 
total stellar mass in ETGs above any given mass, the LF($z$) would have
to change, either in shape or in normalization (to keep constant the number 
of galaxies). This possibility is unlikely because Fig. 1 shows that the
shape and normalization remain the same. One can then safely conclude that 
the majority of the ${\cal M} > 10^{11} M_{\odot}$ ETGs were already in
place at $z\sim 0.8$, have evolved passively since then, and that
virtually no new massive ETGs are formed within this redshift
interval.
This result is not affected by incompleteness effects
(Section 2.1) or by the ``progenitor bias'' (van Dokkum \& Franx 2001),
as the constant number density of luminous ETGs implies that the
progenitors up to $z\sim 0.8$ are all included in the ETG sample.
Near-infrared surveys unveiled galaxies at $z>1.5-2$ with 
properties (star formation rates, masses, metallicity, clustering) 
which qualify them as the progenitors of the massive 
ETGs at $0<z<0.8$, and showed that the star formation activity in 
massive galaxies at $1.4<z<2.5$ is enough to account for all ${\cal
M}>10^{11}$ M$_{\odot}$ red galaxies by $z\sim1$ (Daddi et al 2005b).

$\bullet$ In contrast, at fainter magnitudes the LF is progressively 
depopulated with increasing redshift (Fig. 1 and 3), an effect
that cannot be explained solely in terms on incompleteness (Section
2.1). The progenitor bias appears to be at work here, with an increasing 
fraction of the progenitors to local ETGs which do not already qualify as 
such at higher redshift. A possibility is that these progenitors are still 
star-forming galaxies which are too blue for fulfilling the selection
and which would gradually populate the low-luminosity part of ETG red 
sequence since $z\sim1$ by the progressive quenching of star formation 
in galaxies with masses $<10^{11} M_{\odot}$.

$\bullet$ The relative contribution of this residual star formation to
the build up of the final mass of local ETGs remains to be
assessed. However, the homogeneity of local cluster and field ETGs
(Bernardi et al. 2003; Thomas et al. 2005) argues for the
bulk of the stellar mass having formed at $z>1$ even in low mass ETGs. 
Indeed, even a low level of star formation (e.g. the ``frosting'' scenario 
of Trager et al.  2000) is sufficient to make the color of galaxies too 
blue for qualifying them as color-selected ETGs, whereas quenching such 
residual SF should ensure that the galaxies will quickly join the red 
sequence.

$\bullet$ With the vast majority of ETGs more massive than $\sim
10^{11} M_{\odot}$ being already in place by $z\sim 0.8$, there
appears to be no much room left for any major contribution of dry
merging events in increasing the number of the most massive ETGs.
This appears to be in a agreement with the very low rate
of dry merging events estimated from SDSS data (Masjedi et al. 2005)
and from the most recent results on the stellar mass function
evolution (Caputi et al. 2006; Bundy et al. 2006), but at variance
with other estimates (e.g., van Dokkum 2005; Bell et al. 2006). On the
other hand, dry merging does not change the ETG contribution to the
$B$-band luminosity density $j_{\rm B}$, hence the constancy of
$j_{\rm B}$ since $z\sim 0.8$ (Bell et al. 2004a; Faber et al. 2005)
in spite of the passive dimming of ETGs, requires the transformation
of blue to red galaxies to take place, indeed by the progressive
quenching of residual star formation.

$\bullet$ At each redshift there is a critical mass above which virtually all
ETGs appear to be in place, and the trend of this critical mass with $z$ 
parallels  that of the ``quenching''  mass introduced by Bundy et al. (2006),
as the mass at which the number density of star forming and passive ETGs is 
the same (see Fig. 2).

$\bullet$ All these findings bring new evidence in favor of the
``downsizing'' scenario in galaxy formation which is apparent at low
as well as higher redshifts (see Section 1).
However, our findings extend the concept from the mere {\it
star formation} (stars in more massive galaxies are older) to the
{\it mass assembly itself}, i.e., with massive galaxies being the
first to be assembled (see also Bundy et al. 2006). Downsizing in star
formation may be a natural expectation in a hierarchical galaxy
formation scenario, provided that a suitable mechanism is found to quench
star formation at earlier times in more massive galaxies (e.g., Granato
et al. 2004; De Lucia et al. 2006; Rasera \& Teyssier 2006; Bower et al.
2005). 
However, downsizing in mass assembly seems to be harder to achieve in current 
renditions of the $\Lambda$CDM paradigm, where stars in more massive 
ETGs are indeed older and formed at fairly high redshift, but the most 
massive galaxies are the last (not the first) to be fully assembled, and 
are so mostly by dry merging at $z<1$. This is shown by Fig. 3, where the 
predictions of De Lucia et al. (2006) (dashed line) based on the {\it 
Millennium Simulation} (Springel et al. 2005) show that the brightest, 
most massive ETGs are those which space density most rapidly drops with 
increasing redshift, contrary to the observed evolution of their LF.

$\bullet$ Being more directly connected to the evolution of dark
matter halos, the apparent ``top-down'' {\it mass} assembly of
galaxies should provide a more fundamental test of the $\Lambda$CDM
scenario and its renditions than the mere downsizing in star
formation. As pointed out by Hogg (2005), the empirical dry merging
rate appears to be ``lower, much lower than the rate at which
ETG-hosting dark matter halos merge with one another''.

\begin{acknowledgements}
We thank Sandy Faber, Eric Bell, Toru Yamada, the DEEP2, COMBO-17, and 
SXDS teams for kindly providing their luminosity functions in electronic
form, Lucia Pozzetti and Claudia Scarlata for useful discussion, and 
the anonymous referee for the constructive comments. ED 
gratefully acknowledges NASA support through the Spitzer Fellowship 
Program, award 1268429.
\end{acknowledgements}

\end{document}